\documentclass[sigconf]{acmart}

\usepackage{makecell}

\usepackage{enumitem}

\definecolor{box-bg-yellow}{RGB}{255,254,244}
\definecolor{bar-yellow}{RGB}{255,254,196}

\newsavebox{\obsboxsave}

\newlength{\obsbarwidth}
\newlength{\obsboxpad}
\newenvironment{obscolorbox}[2]{%
  \def\obsbgcolor{#1}%
  \def\obsbarcolor{#2}%
  \par\smallskip
  \noindent
  \begin{lrbox}{\obsboxsave}%
  \begin{minipage}{\dimexpr\linewidth-\obsbarwidth-2\obsboxpad\relax}%
  \ignorespaces
}{%
  \end{minipage}%
  \end{lrbox}%
  \begingroup
  \setlength{\fboxsep}{0pt}%
  \colorbox{\obsbgcolor}{%
    \makebox[\linewidth][l]{%
      {\color{\obsbarcolor}%
       \rule[-\dp\obsboxsave]{\obsbarwidth}{\dimexpr\ht\obsboxsave+\dp\obsboxsave\relax}}%
      \hspace*{\obsboxpad}%
      \usebox{\obsboxsave}%
      \hspace*{\obsboxpad}%
    }%
  }%
  \endgroup
  \par\smallskip
}

\newenvironment{yellowbox}{%
  \begin{obscolorbox}{box-bg-yellow}{bar-yellow}%
}{%
  \end{obscolorbox}%
}

\usepackage{subcaption}

\setcopyright{none}              % remove copyright statement
\renewcommand\footnotetextcopyrightpermission[1]{} % remove first-page copyright footnote
\copyrightyear{2026}
\acmYear{2026}
\setcopyright{cc}
\setcctype{by-nc-nd}
\acmConference[IGSC 2026]{International Green and Sustainable Computing Conference}{June 22--24, 2026}{Canandaigua, NY, USA}
\acmBooktitle{International Green and Sustainable Computing Conference (IGSC 2026), June 22--24, 2026, Canandaigua, NY, USA}
\acmDOI{10.1145/3797248.3815412}
\acmISBN{979-8-4007-2520-3/2026/06}

\begin{document}

\title{Advancing Environmental Sustainability in Data Centers via Carbon Depreciation Models}

% \author{
% #xxxx
% }
\author{Shixin Ji}
\affiliation{%
  \institution{Brown University} 
  \city{Providence}
  \country{USA}
  }
\email{shixin_ji@brown.edu}

\author{Zhuoping Yang}
\affiliation{%
  \institution{Brown University} 
  \city{Providence}
  \country{USA}
  }
\email{zhuoping_yang@brown.edu}

\author{Xingzhen Chen}
\affiliation{%
  \institution{Brown University }
  \city{Providence}
  \country{USA}
  }
\email{xingzhen_chen@brown.edu}

\author{Alex K. Jones}
\affiliation{%
  \institution{Syracuse University}
  \city{Syracuse}
  \country{USA}
  }
\email{akj@syr.edu}

\author{Peipei Zhou}
\affiliation{%
  \institution{Brown University}
  \city{Providence}
  \country{USA}
  }
\email{peipei_zhou@brown.edu}

\begin{abstract}
% Very interesting webpage, though some information here needs to be updated or double-checked from the NVIDIA Official Website
% https://blog.paperspace.com/understanding-tensor-cores/
\label{sec:abs}

Recent improvements in energy efficiency and renewable energy integration have increased the relative importance of embodied carbon in data centers, motivating improved provisioning strategies. 
Conventional approaches primarily minimize operational energy, but this perspective is increasingly insufficient for sustainability.
In this paper, we propose carbon depreciation models to encourage longer hardware lifetimes. 
Carbon depreciation assigns a larger portion of embodied carbon to newly provisioned servers, discouraging unnecessary deployment of new hardware. 
As a result, new servers are provisioned mainly for jobs with strict quality-of-service (QoS) constraints, while older servers, whose embodied carbon has largely been recovered, are used for other workloads.
We further argue that both embodied carbon and operational carbon from server idle time should be recovered during active jobs, encouraging provisioning strategies that maintain high utilization. 
We show that prior carbon accounting strategies can be counterproductive: under a greedy scheduler minimizing carbon under QoS constraints, jobs are priced as 25\% cheaper on new hardware than on older hardware.
In contrast, our approach uses a greedy scheduler that prioritizes older hardware through non-linear carbon depreciation, promoting sustainable provisioning. Experimental results show carbon reductions of 28--57\%, depending on  server lifetime assumptions.
\end{abstract}

\maketitle
\thispagestyle{plain}
\pagestyle{plain}

\section{Introduction}
\begin{figure}[tbp]
    \centering \includegraphics[width=0.80\columnwidth]{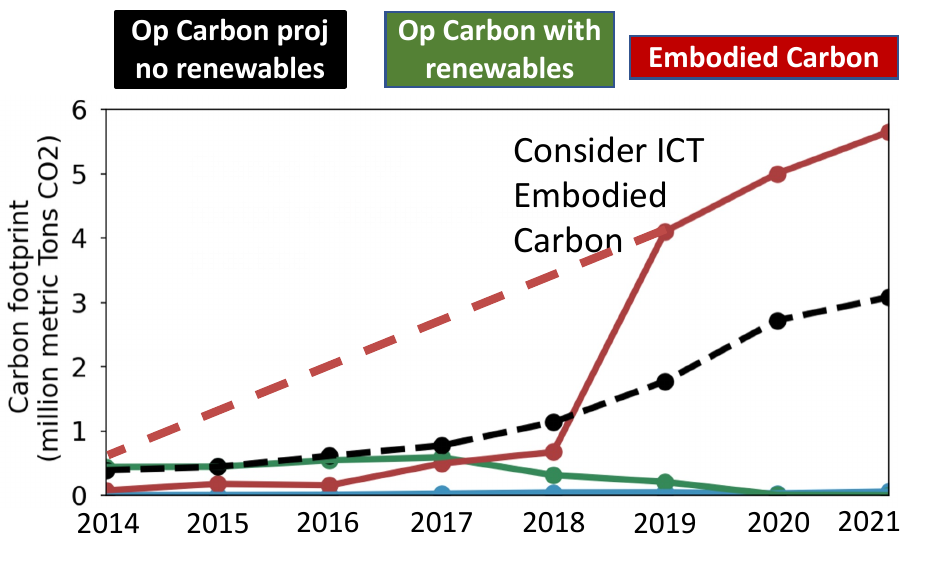}
    \vspace{-15pt}
    \caption{Greenhouse gas emissions from Meta Hyperscalars from 2014--2021.  The initial concern of operational GHG emissions rise (black line) is projected to be mitigated by renewable integration (green line).  Embodied GHG started to be reported in ICT only after 2019 (red line). The reverse projection (dotted red line) shows embodied carbon from ICT may dominate hyperscalar GHG emissions since 2014~\cite{Gupta-Tutorial22}.}
    \label{fig:ghg-hyperscaler-meta}
   \vspace{-15pt}
\end{figure}

\begingroup
\renewcommand\thefootnote{}
\footnotetext{This paper has been accepted at IGSC 2026}
\endgroup
Data centers have become a major source of energy consumption, projected to reach 10\% of global usage within the next 5--10 years~\cite{Electricity}, raising concerns about greenhouse gas (GHG) emissions from powering these facilities.
Mitigation through renewable energy sources~\cite{KWON2020115424} has been actively studied. However, this strong focus on operational decarbonization has, until recently, drawn relatively less attention to the \textit{embodied carbon} of data centers.
Embodied carbon refers to GHG emissions arising from factors outside the direct operation of the data center.
The contribution of Information and Communication Technology (ICT) to embodied carbon has been recognized as substantial and increasing for over a decade~\cite{Jones-Fabrication13}.
Moreover, embodied carbon is now comparable to operational carbon and is likely increasing as operational carbon decreases with greater renewable integration~\cite{Gupta-Tutorial22}, as illustrated in Fig.~\ref{fig:ghg-hyperscaler-meta}.
Therefore, improving the environmental sustainability of data centers requires addressing \textbf{both embodied and operational} carbon contributions of ICT.

Yet, current state-of-the-art techniques that account for embodied carbon, such as Software Carbon Intensity (SCI)~\cite{SCI_project} and Compute Carbon Intensity (CCI)~\cite{switzer2023junkyard}, both propose to combine the operational carbon of a job run on a particular server with the timeslice of the embodied carbon of that machine amortized over its lifetime.  
This approach ignores the age of the server and suggests that the embodied carbon component of a job should be identical whether this is the first or last day of service over that server's lifetime.
This approach has two flaws if the goal is to minimize total carbon: (1) there is no advantage for using older hardware, so all jobs will prefer to run on the newest hardware that best minimizes operational carbon, even if that carbon is relatively small.  
(2) There is no accounting for the \textit{secondary carbon} required for executing jobs.  This secondary carbon comes from the embodied and operational carbon while the server is heavily underutilized.
For users running jobs under these conditions, there is no incentive for users to prefer older hardware, nor to prefer a high utilization of the server.

We propose to address both of these challenges by using carbon job accounting to motivate and better utilize sustainable provisioning practices.  First, we propose to explore using embodied carbon depreciation.  Much like an automobile whose value heavily depreciates early in its lifetime, we propose to explore accounting for significantly more embodied carbon on jobs run earlier in a server lifetime versus later in the lifetime.  
Second, we apply operational and embodied carbon from the server under-utilization (i.e., idle) onto actual jobs run on the server.  
This allows for greedy scheduling strategies to minimize carbon, potentially under the desired quality of service (QoS) constraints in such a way as to provide lower carbon costs when keeping older hardware busy.  
This promotes provisioning strategies to encourage longer lifetimes of ICT hardware and to ensure that these older systems remain highly utilized.   In particular, we make the following contributions:

\begin{itemize}[leftmargin=0pt]
\item We demonstrate that existing methods of CCI/SCI are linear depreciation. 
Moreover, we show that they are inconsistent with increasing server lifetime and do not account for secondary carbon.
\item We explore non-linear depreciation models for ICT and demonstrate how these models encourage the use of older hardware and quantify resulting carbon savings.
\item We show how secondary carbon accounting promotes high server utilization.
\item We conduct a scheduling vs provisioning study under different depreciation and secondary carbon accounting strategies for 
    %\sxj{
    a variety of relevant workloads.
    %}.
\end{itemize}

\begin{figure}[tbp]
    \centering
\includegraphics[width=0.75\linewidth]{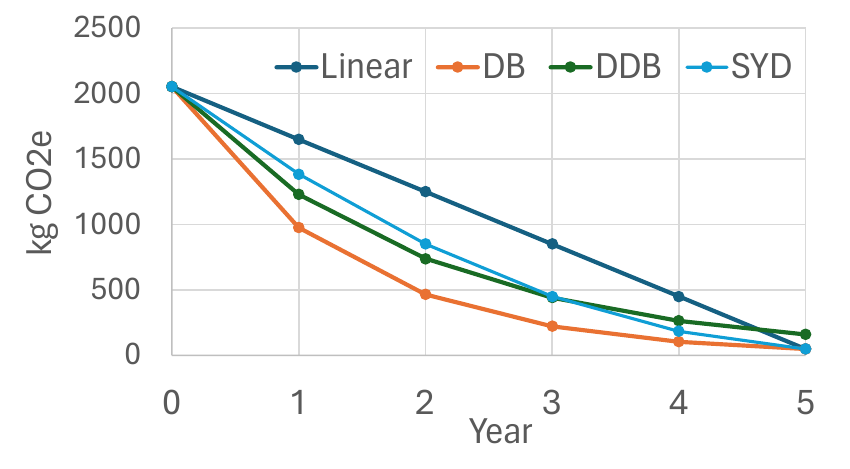}
    \vspace{-17pt}
    \caption{Impact of different depreciation functions.}
    \label{fig:depreciation}
\end{figure}
\begin{table}[tbp]
\renewcommand{\tabcolsep}{.8pt}
\caption{g\,CO$_2$e/s for different carbon depreciation models based on $C_{em}=2050\,\mathrm{kg\,CO}_2\,\mathrm{e}$ with a five year lifetime $L=5$.}
\label{tab:carbon-per-time}
\begin{tabular}{c|rrrr||rrrr||rrrr}
\hline\hline
\textbf{Year} & \multicolumn{1}{c}{\textbf{Lin}} & \multicolumn{1}{c}{\textbf{DB}} & \multicolumn{1}{c}{\textbf{DDB}} & \multicolumn{1}{c||}{\textbf{SYD}} & \multicolumn{1}{c}{\textbf{Lin}} & \multicolumn{1}{c}{\textbf{DB}} & \multicolumn{1}{c}{\textbf{DDB}} & \multicolumn{1}{c||}{\textbf{SYD}} & \multicolumn{1}{c}{\textbf{Lin}} & \multicolumn{1}{c}{\textbf{DB}} & \multicolumn{1}{c}{\textbf{DDB}} & \multicolumn{1}{c}{\textbf{SYD}} \\
& \multicolumn{4}{c||}{Utilization=1} & \multicolumn{4}{c||}{Utilization=0.8} & \multicolumn{4}{c}{Utilization=0.3}\\\hline
1 & 12.7 & 34.1 & 26.0 & 21.1 & 15.9 & 42.6 & 32.5 & 19.9 & 42.3 & 113.5 & 86.7 & 70.5\\
2 & 12.7 & 16.2 & 15.6 & 16.9 & 15.9 & 20.3 & 19.5 & 21.1 & 42.3 & 54.0 & 52.0 & 56.3 \\
3 & 12.7 & 7.7  & 9.4  & 12.7 & 15.9 & 9.6 & 11.8 & 15.9 & 42.3 & 25.7 & 31.2 & 42.3 \\
4 & 12.7 & 3.7  & 5.6  & 8.5 & 15.9 & 4.6 & 7.0 & 10.6 & 42.3 & 12.2 & 18.7 & 28.2 \\
5 & 12.7 & 1.8  & 3.4  & 4.2 & 15.9 & 2.3 & 4.3 & 5.3 & 42.3 & 5.8 & 11.2 & 14.1  \\ \hline\hline
\end{tabular}
\end{table}
\renewcommand{\tabcolsep}{6pt}

\section{Motivation}
\label{sec:motivation}

The trends for embodied carbon of server products are relatively similar for similarly configured systems across generations~\cite{isvlsi24scarif}.  Thus, provisioning new systems immediately introduces a significant source of carbon (embodied carbon) before the system has completed any useful work.
Yet server lifetimes cannot be extended indefinitely, and new workloads may demand the capabilities of newer server generations to meet QoS requirements.
% In this section, we offer a motivational discussion of why new models of combining embodied carbon with data center jobs are important to support sustainable provisioning.

% \subsection{Exploring Carbon Depreciation} 
% \label{sec:motivation-depreciation}
Previous approaches~\cite{switzer2023junkyard,SCI_project} suggest that embodied carbon should be amortized and assigned to data center jobs regardless of equipment age, i.e., independent of the location on the lifetime curve of the equipment when satisfying the request.
Thus, carbon for a particular job $C_i$ is a function of operational carbon $C_{i,op}$ and a portion $\delta$ of embodied carbon $C_{em}$, or $\delta C_{em}$. $\delta = t_i/L$, where $t_i$ is the time the job uses the resource and $L$ is the lifetime of the resource.

However, it is well known that technologies like ICT are particularly sensitive to age.  For instance, trends like Moore's law have created an expectation that ICT should have an extremely short lifetime as it will become obsolete in short time periods.  
To quantify the value of these technologies, it is common to use an economic principle of depreciation.
Thus, in the current  modeling~\cite{switzer2023junkyard,SCI_project} $C_i = C_{i,op} + \delta C_{em}$.  Fig.~\ref{fig:depreciation} shows this as a \textit{linear depreciation} model, sometimes referred to as straight-line depreciation~\cite{PrinAccounting11} (blue) for a server with $C_{em}= 2050\,\mathrm{kg\,CO}_2\,\mathrm{e}$.  In this model, the $C_{em}$ is spread equally over $L$ such that $\delta$ for job $i$ is the same on the first or last day the equipment is in service.  By the end of the $L=5$ years lifetime, $2000\,\mathrm{kg\,CO}_2\,\mathrm{e}$ has been depreciated.  

In economics there are other depreciation formulas designed to reflect the age of the equipment.
The \textit{declining balance} (DB) depreciation formula~\cite{PrinAccounting11}, shown in orange in Fig.~\ref{fig:depreciation}, pushes much more of the $C_{em}$ to the early years of the lifetime.  During year 1, linear depreciation recovers only $400\,\mathrm{kg\,CO}_2\,\mathrm{e}$ while DB recovers $820\,\mathrm{kg\,CO}_2\,\mathrm{e}$, but only $106\,\mathrm{kg\,CO}_2\,\mathrm{e}$ in the last year.  Now $\delta = F(A,C_{em},L)$ where $A$ is the ICT age.  Fig.~\ref{fig:depreciation} also shows \textit{double declining balance} (DDB) and sum of years difference (SYD) depreciation~\cite{PrinAccounting11}, which are still based on age, with each having less aggressive carbon recovery in early years and more aggressive carbon recovery in later years.
To illustrate this, the equivalent CO$_2$ from Fig.~\ref{fig:depreciation} is shown per second in Table~\ref{tab:carbon-per-time}.  A 1 second job run on a fully utilized machine in year 1 or 5 is $12.7\,\mathrm{g\,CO}_2\,\mathrm{e}$ of embodied carbon, but in DB model is nearly $3\times$ larger in year 1 and almost $10\times$ smaller in year 5.  
% Thus, with a linear depreciation model, any savings in operational carbon from energy efficiency (including performance benefit) with a new machine will make that machine more desirable in sustainability.  
% However, with DB, DDB, and SYD models, newer machines' embodied carbon will be higher in year 1 than year 5, with a ratio that is highest with DB and most moderated with SYD.  Thus, the carbon optimization will be a function of both $C_{em}$ and $C_{op}$.

Moreover, just because a server is idle, it does not mean that the embodied (and operational) carbon can be ignored.  In Table~\ref{tab:carbon-per-time} we show how the embodied carbon scales for utilization $U \in \{1,0.8,0.3\}$ to show $C_{em}$ .
% for the lower bound, a relatively high utilized, and a relatively low utilized server, respectively.  
This motivates keeping servers highly utilized.  Morepver, using a non-linear depreciation model, even with lower utilizations of older servers, the embodied carbon can be much lower than that of a highly utilized new server. 

\section{Related Work}
\label{sec:related}
Academic research efforts have made considerable progress over the last decade to estimate or profile the hardware systems' carbon cost and use the profiling information to guide the design toward sustainable systems.
Earlier works in the 2000s focused on the operation carbon cost of computing systems~\cite{fan2007power,barroso2007case,rajamani2003evaluating_request_distribution_schemes}.
% , which is consistent with industry emphasis on reducing energy consumption, provided significant cost savings as data centers started to become prevalent.  
In more recent years, 
% with seminal work starting about a decade ago
~\cite{Jones-Fabrication13} noting that the embodied costs are consuming more and more of the overall carbon footprints in a product's lifetime.
% consequently requiring more attention.
% Life Cycle Assessment (LCA) is a common method to quantitatively evaluate the GHG emissions of computing systems throughout their whole lifetime.
% Other than the LCA tools, more carbon cost estimation and analysis approaches are proposed to derive insights into sustainable computing systems.
Greenchip~\cite{igsc2016greenchip,KLINE2019322} is the earliest predictive estimation tool to comprehensively understand the environmental impact of computing systems.
ACT~\cite{ACT} is a system modeling tool that is built, like Greenchip, with lower-level data from industry fabs.
More recently, ECO-CHIP~\cite{sudarshan2024eco} has been proposed to estimate the carbon cost of chiplet-based architectures.
U-DUCT~\cite{u_duct} and CarbonClarity~\cite{CarbonClarity} address the uncertainty in the estimation.
Based on these carbon cost estimation tools, more studies are being conducted to gain insights into more specific domains.
\cite{li2023toward} integrates the modeling of embodied and operational carbon to understand the environmental impact of high-performance computing (HPC) systems. 
FOCAL~\cite{eeckhout2024focal} estimates the carbon cost efficiency of a series of different processor mechanisms.
LLM carbon~\cite{faiz2023llmcarbon} and LLMCO2~\cite{llmco2} estimate the carbon cost in training and inferencing large language models.
and~\cite{sustainable_carbon_aware} proposes carbon-aware scheduling of LLM in data centers.
% GreenScale~\cite{kim2023greenscale}
% proposes an application scheduling mechanism, utilizing the various carbon intensities in different locations to reach a lower overall operational carbon cost.  
Fair-CO2~\cite{fairco2} proposes to fairly attribute carbon cost in data centers via game theory.
REFRESH~\cite{zhou2023refresh}
poses the opportunity to save carbon by building new FPGA chiplet systems by reusing the decommissioned old FPGA chips and extending their lifetime.

\section{Methodology}
\label{sec:method}
The manufacturing and operational phases of ICT hardware generally contribute more than 95\% of the carbon costs during their lifetime~\cite{ACT,igsc2016greenchip,KLINE2019322}.  Thus, we expand on our methodology for quantifying the environmental impact of computing systems by quantifying these two components of carbon production.  Recall, we consider carbon $C$ as $C=C_{op}+C_{em}$.
For per-job carbon cost, we quantify the carbon cost for a specific task $i$ as $C_i$ and its embodied and operational components as $C_{i,em}$, and $C_{i,op}$.
\subsection{Embodied Carbon Modeling}
To effectively consider embodied carbon in data center carbon minimization requires a methodology to estimate embodied carbon and a technique to apply an appropriate portion of it to jobs on that server and/or related ICT.
% \subsubsection{Quantifying system embodied carbon} 
% Currently, mainstream carbon cost estimation tools like ACT~\cite{ACT} and Greenchip~\cite{KLINE2019322,igsc2016greenchip} usually apply a bottom-up method and generate much smaller results than hardware vendor reports. 
% Some works also tried to use the LCA reports directly~\cite{acun2023CarbonExplorer}. However, these reports are specific to one setup and do not include accelerators like GPUs.
We use a data-driven, end-to-end tool, SCARIF~\cite{isvlsi24scarif}, which is flexible to different server configs and is able estimate the environmental impact of accelerator, to estimate the total embodied carbon costs of different server generations.

% \subsubsection{Accounting for the per-job embodied carbon using depreciation models}
\label{sec:recover-carbon}
As shown in Fig.~\ref{fig:depreciation}, with different kinds of depreciation models, the total embodied carbon cost of ICT hardware could be distributed over each year of the lifetime.
We assume this yearly embodied cost is evenly distributed over every second of the year.
Thus, when the secondary carbon is not accounted for, the embodied carbon cost of a task is the cost in the task time:
\begin{equation}
\begin{aligned}
    C_{i,em} &= \delta C_{em} \quad ; \quad
    \delta = {t_{task}\over t_{year}}F(A, C_{em}, L)
    %F &\in \{Linear, DB, DDB, SVD \}
\end{aligned}
\end{equation}
When the secondary carbon is accounted for, each task would share all the embodied costs generated in this year.
Thus, the carbon cost will be related to the utilization $U$ of the server.
The modeling will change as follows, where $J$ stands for the total number of jobs:% 
\begin{equation}
\begin{aligned}
    \delta^\prime &= {1\over J}F(A, C_{em}, L) \quad ; \quad
    J = {{t_{year} \cdot U}\over t_{task}}
\end{aligned}
\end{equation}
\subsection{Operational Carbon Cost Modeling}
The carbon costs generated in the operational phase of ICT hardware mainly come from energy consumption. Here, we model the operational cost of a system by its energy consumption $E$ and the carbon intensity $CI$ of the local grid, i.e., $C_{op} = E \cdot CI$.

\subsubsection{Modeling carbon intensity}
The $CI$ (in terms of kgCO$_2$e$/$kWh) stands for how much carbon is needed to generate a unit of energy, in the form of electricity, to operate the data center, dominated by the ICT.
The $CI$ value highly depends on the electrical grid where the data center is located and is related to the ratio of renewable energy in the grid, and varies frequently and greatly with time.

To quantify the geographical differences in carbon intensity, we use the grid mix concept reported in prior work~\cite{ollivier2022sustainable}.
% For the temporal differences of the carbon intensity, \cite{sukprasert2024implications,maji2024untangling} use  \textit{average carbon intensity} to represent the weighted average from all energy sources in the electrical grid to satisfy the energy resource. 
% Since the change of power generators usually takes a long period, the \textit{average carbon intensity} within a year can be regarded as stable.
% Moreover, the variance of $CI$ within a day could also be averaged to a constant when the pattern of workloads is also fixed.
% For advanced scheduling techniques like
% \cite{souza2023casper,lechowicz2023online}, their benefits could also be represented by a smaller average carbon intensity.
The variance of $CI$ within a day, and advanced scheduling techniques like \cite{souza2023casper}, can be represented by shifting the average $CI$.
the absolute amount of energy consumption will also affect the actual carbon intensity, which has been discussed by~\cite{sukprasert2024implications, maji2024green}.
% This is due to the working pattern of the electrical grid:
% This is because when energy consumption is small, it can be satisfied by renewable energy.
% But when it is large, the carbon intensity will be the mix of renewable and non-renewable sources and be much higher.
We use two carbon intensity values, $CI_a$ and $CI_i$, to represent the average carbon intensity at the active and idle state of a server. 

\subsubsection{Computing per-job operational carbon cost}
When secondary carbon is not considered, the per-job carbon cost will be:
When secondary carbon is not considered, the per-job carbon cost will be:
% \akj{be careful not to be repetitive, did you already say this?}
\begin{equation}
\begin{aligned}
    C_{i,op}&=E_a\cdot CI_a \quad ; \quad
    E_a =P_a \cdot t_{task}
\end{aligned}
\end{equation}
Here, the $P_a$ is the server's power when active.
Then, when the secondary carbon is considered, besides the active state, the server will also have a much smaller power $P_i$ in the idle state, and all carbon costs generated within one year will be equally distributed to each task in this year:
\begin{equation}
\begin{aligned}
    C_{i,op}^\prime &=E_a\cdot CI_a + E_i\cdot CI_i \quad ; \quad
    E_i = {{t_{year} \cdot (1-U) \cdot P_i}\over{J}}
\end{aligned}
\end{equation}

\begin{table}[h]
    \centering
    \footnotesize
    \caption{
    Details of three studied systems. CPU Cores is the number of allocated or used CPU cores in the system.
    %The CPU/GPU ratio is set to close to 8 threads per GPU. 
    %CPU number is the ratio of the CPU threads (e.g., allocated vCPUs in public data center, or actual CPU in local data center) of the total CPU resources within a chip \akj{explain *}
    %.
    % Setups of systems, \sxj{need to be improved}: (1) now a system has 4 GPUs, (2) add SCARIF reported values
    %Details of the four studied server systems. Three are public datacenter servers and one is a local datacenter server. CPU number is the ratio of the CPU threads (e.g., allocated vCPUs in public data center, or actual CPU in local data center) of the total CPU resources within a chip.
    %Each vCPU is a thread of either an Intel Xeon core or an AMD EPYC core.)
    }
    \begin{tabular}{cccccc}
        \hline
        System & CPU & \makecell{CPU \\\#Cores} & GPU & \makecell{GPU \\Num} &\makecell{Total Embodied\\ Cost (kg\,CO$_2$e)} \\
        \hline
        % \hline
        
         1   & \makecell{Intel Xeon \\E5-2686 v4}& 32 & \makecell{NVIDIA\\V100} & 4 &655.18\\
        \hline
         2  & \makecell{AMD \\EPYC 7R32} & 32 &  \makecell{NVIDIA\\A10G} & 4 &595.5\\
        \hline
         3  & \makecell{Intel Xeon\\Gold 6346} & 32 & \makecell{NVIDIA \\RTX5000Ada} & 4 &925.94 \\
        \hline
    \end{tabular}
    \label{tab:studied_systems}
\end{table}

% \begin{table}
%     \centering
%     \footnotesize
%     \caption{
%     Setups of systems, \sxj{need to be improved}: (1) now a system has 4 GPUs, (2) add SCARIF reported values
% \begin{tabular}{cccccc}
%         \hline
%         No. & Server & CPU & \makecell{CPU\\Num} & GPU & \makecell{GPU \\Num}  \\
%         \hline
%         \hline
        
%         1 &\makecell{AWS\\ p3.2xlarge}  & \makecell{Xeon E5\\-2686 v4}& 0.22* & \makecell{NVIDIA\\V100} &1 \\
%         \hline
%          2 &\makecell{AWS\\G5.2xlarge} & \makecell{AMD \\EPYC 7R32} & 0.083*&  \makecell{NVIDIA\\A10G} & 1 \\
%         \hline
%          3 &\makecell{AWS\\G5.16xlarge} &\makecell{AMD \\EPYC 7R32}  & 0.67*& \makecell{NVIDIA\\A10G}&1 \\
%          \hline
%          4 &Local & \makecell{Intel Xeon\\Gold 6346} & 1 & \makecell{NVIDIA \\RTX5000Ada} & 1 \\
%         \hline
%     \end{tabular}
%     \label{tab:studied_systems}
% \end{table}

% \input{tables/AppParam}

\subsection{Application Benchmarking}
% In this work, we profile the actual performance of 10 different representative applications on 3 generations of GPU devices to determine their execution characteristics.
We use three GPUs: Nvidia V100 (2017), A10G (2020), and RTX5000 Ada (2023).
They are released in the same time gap (three years) and have a similar thermal design power (TDP).
In this work, we study the system setups as shown in Table~\ref{tab:studied_systems} for discussion. 
We pair the type of GPU with the CPU that was released in similar years.
throughput, 99\% latency, and power in active and idle state are collected for GPUs.
% When measuring the workloads, we collect the throughput, 99\% latency, and power in the active state of each GPU. 
% % We also collect the GPU power in the idle state.
For the CPUs, our data shows that their utilization is less than 10\% in the active state, thus their power is not included.
We utilize the NVIDIA Triton~\cite{nvidia_triton} framework to execute AI inference workloads.
To achieve high performance, we leverage the Triton model analyzer to search for the optimal configuration. 
We study nine representative models including computer vision (CV) models (DenseNet, ResNet-50, ResNet-101, VGG-19), natural language processing(NLP) models (Bert-B, Bert-L) , and large language models (LLMs) (OPT-2.7B, Falcon-7B, Persimmon-8B).
% , as shown in Table~\ref{tbl:applicationModel}.

\begin{figure}[tb]
\centering
\includegraphics[width=0.75\columnwidth]{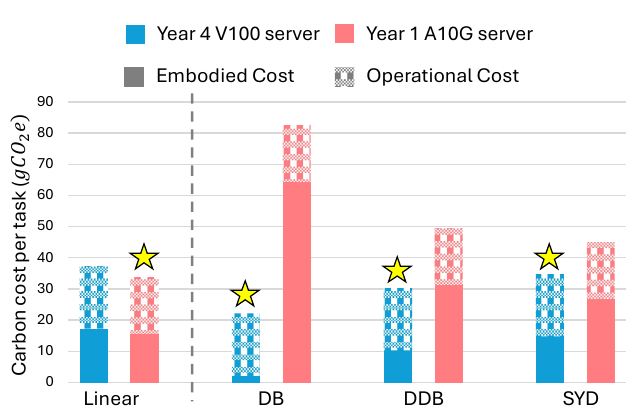}
\vspace{-15pt}
\caption{Carbon cost comparison in 2020 between year 4 V100 server and year 1 A10G server under 4 different depreciation models. The required throughput is set to 200 infer/s.} 
\label{fig:depreciation_model_overview}
\vspace{-15pt}
\end{figure}

\begin{figure}[tb]
\centering
\includegraphics[width=0.75\columnwidth]{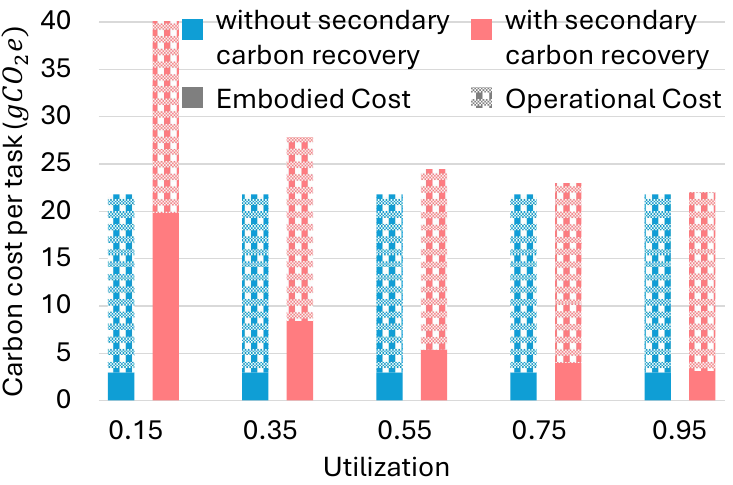}
\vspace{-14pt}
\caption{Carbon cost comparison between estimations w/o \& w/t considering secondary carbon for a year 4 V100 server. 
}
\label{fig:idle_state_comparsion}
\vspace{-18pt}
\end{figure}

\section{Analysis about Impacts of Different Carbon Cost Modeling Setups}
\label{sec:carbon_analysis}

To demonstrate the methodology, we first study the scenario of provisioning GPU servers in 2020. 
In this scenario, the provisioner needs to choose to continue using a 4-year-old V100 server (released in 2017) or upgrade to the brand new A10G server (released in 2020).
We study this choice using the ResNet50 workload.  The estimated lifetime for both servers is 6 years for carbon embedded cost depreciation. %\akj{how can it be 6 years if you're not sure you're replacing the V100?}
The $CI_a$ and $CI_i$ are set to 0.188 and 0.019 kgCO$_2$e$/$kWh.
We will discuss the generality of our methodology and evaluate other workloads in Section~\ref{sec:sensitivity-study}.

\subsection{Criterion 1: Choice of Depreciation Models}
Fig.~\ref{fig:depreciation_model_overview} shows the carbon cost per 1 million inferences of ResNet 50 executed on the V100 and A10G servers using four different depreciation models.
In this comparison, the requested throughput of each server is 200 inferences/s. 
 This corresponds to a utilization of 0.17 and 0.12 for the V100 and A10G servers.
%linear 
A carbon-minimizing approach will prefer the A10G-based server with a linear carbon depreciation model, as it has a lower $\delta C_{em}$ and lower $C_{i, op}$.
% For other depreciation models, the carbon minimizing approach will prefer the V100 server, as shown in Fig.~\ref{fig:depreciation_model_overview}.
% Using different depreciation models will not affect the operational carbon portion of the job carbon cost.  Thus, the A10G server still has a lower operational cost per job.

For other depreciation models, the carbon minimizing approach will prefer the V100 server, as shown in Fig.~\ref{fig:depreciation_model_overview}.
Using different depreciation models will not affect the operational carbon portion of the job carbon cost.
However, the depreciation models of DB, DDB, and SVD tend to assign more carbon cost to the early years and less to the later years.
So, using brand new A10G servers becomes expensive, and the effect of embodied carbon cost easily overwhelms the effect of operational carbon. 
For example, when the DB model is used for a job, the embodied carbon cost for running the job on the A10G server is 64\,g\,CO$_2$e,
which is $4.14\times$ of that under the linear model.
On the other hand, the embodied cost of the job is only 2.2\,g\,CO$_2$e for running on the V100 server, only 0.13$\times$ under the linear model.
The DDB and SVD depreciation models follow the same pattern but are less aggressive compared to the DB model, with more carbon remaining in the later years.
\begin{yellowbox}
{\textbf{Observation.}} 
The traditional linear depreciation model attaches importance to the operational carbon cost and thus prefers faster and newer devices. 
This is not compatible with the principle of reusing. On the other hand, using the DB, DDB, and SVD depreciation models will address the importance of embodied carbon costs, encourage the extending the lifetime of hardware, and avoiding over-provisioning.
\end{yellowbox}

\subsection{Criterion 2: Secondary Carbon}

Fig.~\ref{fig:idle_state_comparsion} shows the carbon cost \textit{accounted per task} of the same V100 server in the year 2020 between estimations without and with secondary carbon.\footnote{This is similar to prior work that has noted a relationship between carbon efficiency and utilization of dark silicon~\cite{Dark-Silicon-Harmful} and GPU~\cite{wu2022sustainable} accelerators.  However, unlike prior work, we explicitly differentiate between carbon due to computing a task (primary carbon) vs. carbon assigned to a task due to poor utilization (secondary carbon). }
The linear depreciation model is used.
For estimations without secondary carbon, the carbon cost will remain the same for different throughputs.
One disadvantage of this method is that the summation of the per-job carbon cost is not equal to the total carbon cost per year. 
The ignored part, i.e., the secondary carbon cost, is not trivial.
For example, when throughput is set to 200 (about 20\% Utilization), the carbon cost per task without secondary carbon is 21.8\,g\,CO$_2$e. 
However, if the secondary carbon is considered, it is 37.4\,g\,CO$_2$e.
The carbon cost generated in the idle state, is 41.7\% of the total cost and
71.7\% of the active state.
This high secondary carbon is mainly due to embodied costs, as most embodied carbon is allocated to the idle state.
As the utilization rises, the carbon cost decreases quickly.
This is because (1) the period of the idle state is shortened, and (2) there are more tasks to distribute the secondary carbon cost.

\begin{figure}[tb]
\centering
\includegraphics[width=0.75\columnwidth]{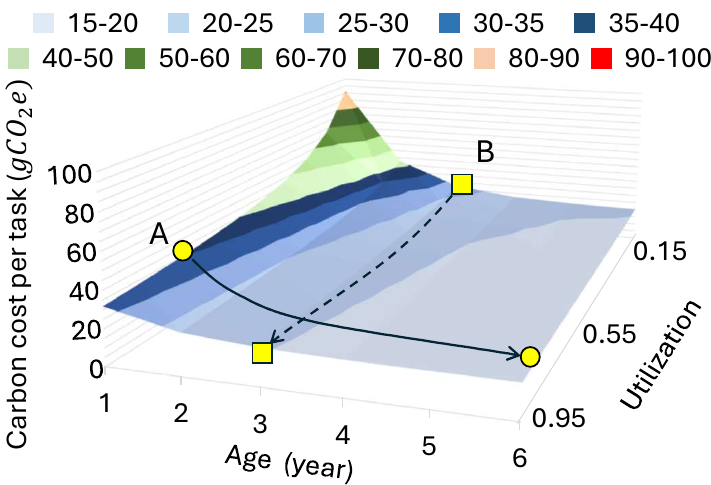}
\vspace{-14pt}
\caption{Per-job Carbon cost of the V100 server in different years and server utilizations. Two trendlines: (A) in different years; and (B) under different utilizations have been marked.}
\label{fig:depreciation_idle_combination}
\vspace{-16pt}
\end{figure}

\begin{figure*}[tb]
%\centering
\begin{subfigure}[t]{0.31\textwidth}
\includegraphics[width=\columnwidth]{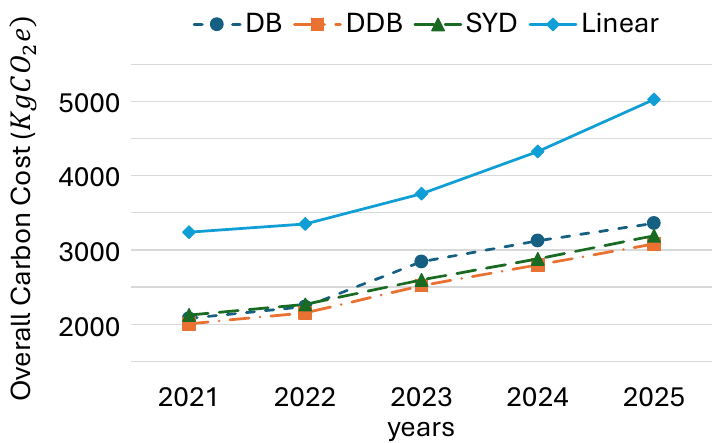}
\vspace{-15pt}
\caption{Provisioning lifetimes match scheduling
}
\label{fig:different_lifetime}
\end{subfigure}%
~
\begin{subfigure}[t]{0.31\textwidth}
\includegraphics[width=\columnwidth]{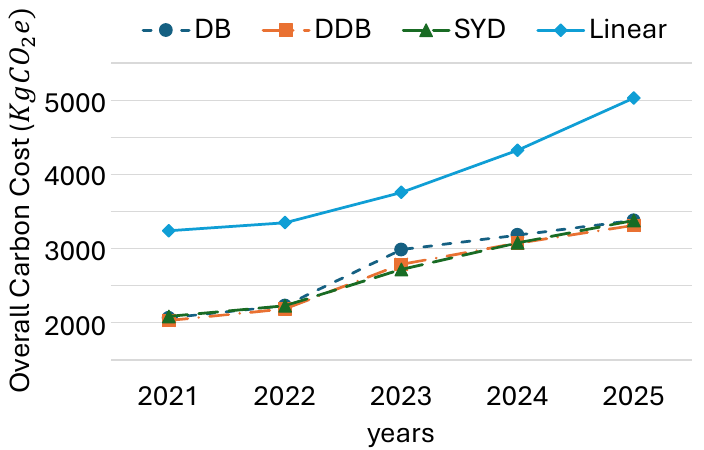}
\vspace{-15pt}
\caption{Lifetimes set 3y, scheduler can extend}
\label{fig:3_year_lifetime}
\end{subfigure}
~
\begin{subfigure}[t]{0.31\textwidth}
\includegraphics[width=\columnwidth]{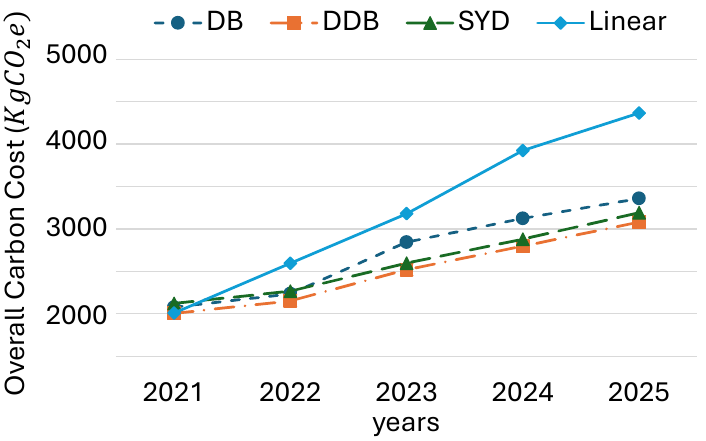}
\vspace{-15pt}
\caption{
Lifetimes set 6y, scheduler can reduce}
\label{fig:6_year_lifetime}
\end{subfigure}

\vspace{-10pt}
\caption{Comparing different depreciation strategies under different lifetime assumptions}
\vspace{-15pt}
\label{fig:linear-vs-nonlinear-for-different-lifetime-constraints}

\end{figure*}

\begin{yellowbox}
{\textbf{Observation.}} 
The secondary carbon should be considered when we distribute the carbon cost to each task. The reasons are: (1) considering the secondary carbon will give a more comprehensive view of the carbon costs generated in the idle state; (2) by considering the secondary carbon, the carbon cost metric could encourage keeping a high utilization of servers and punish over-provisioning.
\end{yellowbox}

Fig.~\ref{fig:depreciation_idle_combination} shows the per-job carbon cost of a V100 server with two critical variables: the depreciation period(age) of the server, and the utilization of the server. 
The DB model is used, and the secondary carbon is considered.
Other server configurations will also follow a similar trend.
As shown in this figure, our methodology will have three impacts on the provisioning strategies as follows:

\textbf{Extending devices' lifetime is encouraged}: Trendline A in Fig.~\ref{fig:depreciation_idle_combination} shows what happens to a server as it ages: In the first year, carbon cost is high, then as the carbon depreciates, the per-job carbon cost drops quickly.
Finally, in the later years of one's lifetime, the old server enjoys a very low carbon cost.
As a result, extending the lifetime of old servers becomes the optimal choice in most cases.

\textbf{Keeping high utilization is desirable}: Trendline B shows the carbon cost of a server as utilization varies. 
As the utilization increases, the carbon cost decreases.
This pushes provisioning fewer servers to keep a high overall utilization of the data center.

\textbf{Over-provisioning of ICT is punished}: 
When upgrading an old device:
(1) a new device has a much higher per-job embodied cost, especially in its first year;
(2) since the new devices are usually faster, the utilization decreases when the workload remains unchanged.
As a result, upgrading will almost always lead to a higher carbon cost.
Then the optimal choice is to reuse the old ones, unless under certain constraints, like the QoS requirements.

\subsection{Overall Carbon Costs: Case Study}
% \begin{violet-box}
% {\textbf{RQ 5.}} \textit{How does the job scheduler impact the provisioning and what is the resulting trend on global-warming potential?}
% \end{violet-box}

Based on the prior takeaway, we quantify the potential for improvement. Figure~\ref{fig:linear-vs-nonlinear-for-different-lifetime-constraints} shows the global warming potential (GWP) in kg\,CO$_2$e under different lifetime assumptions for server configurations using hardware available from 2017-2025 and evaluated over 2021-2025\footnote{Data center server mixes include servers late in their lifetime; since A10G servers replaced V100 servers in 2019, a 2021 mix with three hardware generations is reasonable for a 6-year server lifetime.}.
We also assume a 25\% yearly increase in requested jobs. Because the greedy per-job scheduler under a linear carbon depreciation model prefers newer hardware to minimize GWP, we allowed it to influence the provisioner to replace machines more quickly (3-year lifetimes). This maximized the number of jobs receiving the lowest possible GWP. In contrast, under non-linear depreciation, the scheduler prefers older hardware when QoS is satisfied, favoring 6-year lifetimes. In this case, hardware is replaced only when required by the increasing workload.
When lifetimes match the scheduling preference (Figure~\ref{fig:different_lifetime}), the linear approach substantially increases embodied carbon, making it highly problematic for GWP.  

When we allowed the scheduler to influence provisioning, rather than applying a static replacement approach, and lifetimes were presumed to be short (Figure~\ref{fig:3_year_lifetime}), keeping servers in service longer after their embodied carbon had been recouped tended to best match the oracle (Figure~\ref{fig:different_lifetime}). However, presuming a longer lifetime and replacing servers early not only favored the non-linear models but also showed a separation trend over time (Figure~\ref{fig:6_year_lifetime}). Depending on the policy selected to maximize the number of jobs that run with minimum carbon for each depreciation model, the DDB method saves 7145 (56.9\%), 6308 (47.1\%), and 3521 (28.1\%) kg\,CO$_2$e for the scenarios shown in Figures~\ref{fig:different_lifetime}--\ref{fig:6_year_lifetime}, respectively, over the period from 2021--2025. DDB also performs best among the non-linear depreciation models, saving 1082 (8.6\%), 452 (3.4\%), and 1082 (8.6\%) kg\,CO$_2$e for the same scenarios.

\begin{yellowbox}
{\textbf{Takeaway.}} 
The linear depreciation approach with a greedy scheduler to minimize per-job carbon promotes faster machine replacement and increases overall carbon, which could be considered an instance of Jevon's paradox~\cite{jevon} of efficiency actually increasing consumption.  In contrast using a non-linear distribution of embodied carbon can encourage the scheduler to prefer older hardware, which can actually increase conservancy.  %Moreover, DDB performs better than DB and SYN depreciation models~\cite{PrinAccounting11}.
\end{yellowbox}

% \begin{figure}[tb]
% \centering
% \includegraphics[width=0.8\columnwidth]{figures/sensitivity_study_fig1_v3_croped.pdf}
% \caption{
% Sensitivity analysis of per-job cost ratio of year 1 A10G vs. year 4 V100 varied over grid mix and application. Different grid mixes produce different carbon intensities. Different applications exhibit different energy efficiency gains from A10G over V100.
% }
% \label{fig:sensitivity_fig1}
% \end{figure}

\begin{figure*}[tb]
\centering
\includegraphics[width=0.75\textwidth]{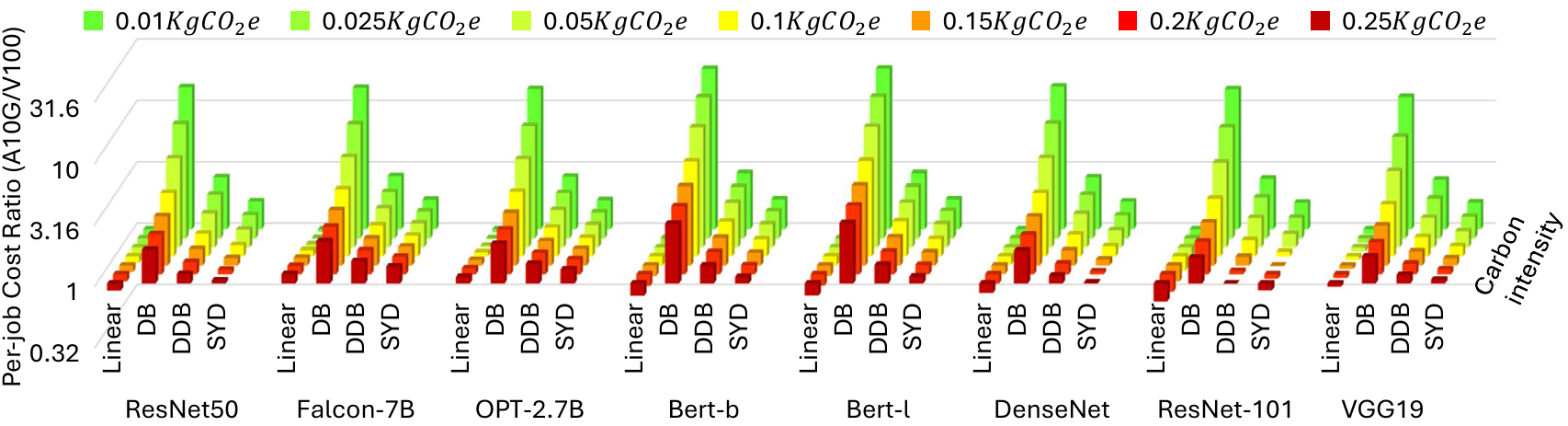}
\vspace{-13pt}
\caption{
Per-job cost ratio of year 1 A10G vs. year
4 V100 in 2020 varied over grid mix (carbon intensities) and depreciation models for different applications. 
}
\label{fig:multi_app_fig1}
\vspace{-14pt}
\end{figure*}

\begin{figure*}[tb]
\centering
\includegraphics[width=0.75\textwidth]{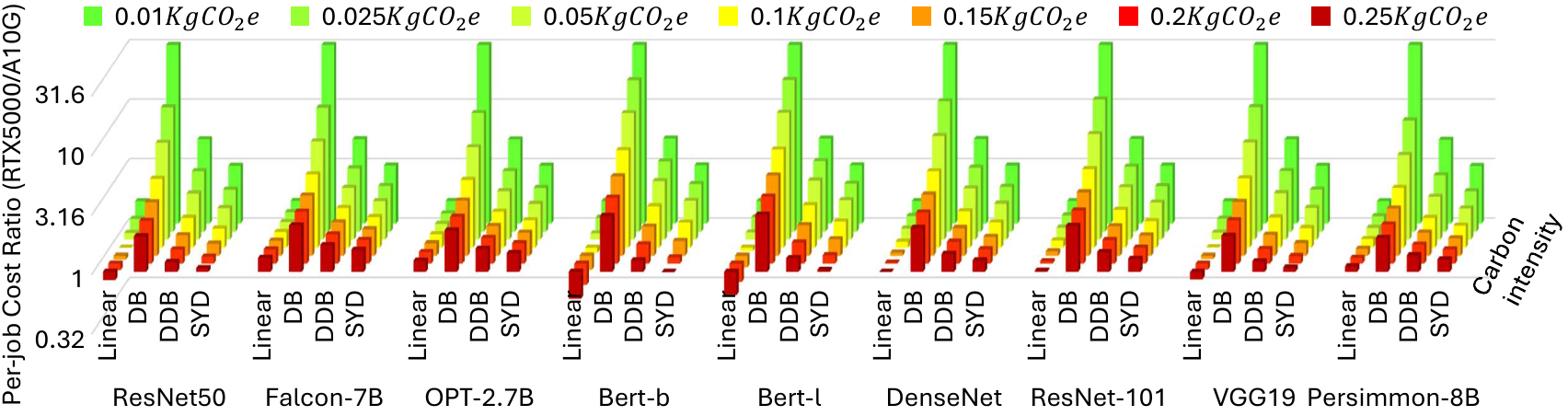}
\vspace{-15pt}
\caption{
Per-job cost ratio of year 1 RTX5000 vs. year
4 A10G in 2023 varied over carbon intensities and depreciation models.
}
\label{fig:multi_app_fig2}
\vspace{-15pt}
\end{figure*}

\section{Sensitivity Study}
\label{sec:sensitivity-study}

In this section, we discuss how two factors, carbon intensity and energy efficiency, affect the design choices.
Then we show the experiment results on different workloads to show the generality.
% \begin{figure}[tb]
% \centering
% \includegraphics[width=0.8\columnwidth]{figures/sensitivity_study_fig1_v3_croped.pdf}
% \caption{
% Sensitivity analysis of per-job cost ratio of year 1 A10G vs. year 4 V100 server.
% }
% \label{fig:sensitivity_fig1}
% \end{figure}

% \subsection{Factors Affecting Design Choices}

In comparing the environmental impact of the old and new devices, the embodied carbon cost is relatively fixed and usually only related to the device model.
However, there are lots of factors affecting the operational carbon cost.
Based on the modeling that $C_{op}=E\cdot CI$, we could divide these factors into two categories: factors affecting the carbon intensity or energy consumption per-job.
Newer servers usually have a gain in the per-job energy consumption due to lower latency or higher energy efficiency, and the carbon intensity determines how largely this translates to carbon cost gain.

% We conduct the sensitivity study of the above two factors in Fig.~\ref{fig:sensitivity_fig1}. 
% This study is based on the setup shown in Section~\ref{sec:carbon_analysis} between the year 4 V100 server and the year 1 A10G server running ResNet-50 workloads.
% The DB depreciation model is used.
% Originally, the A10G server had a 1.47$\times$ higher throughput and a 1.16$\times$ higher energy efficiency.
% We assume that there is a hypothetical software optimization technique for the A10G server, which can increase the throughput to up to 2.5$\times$ V100 throughput while having nearly no impact on the power.
% We use the ratio of per-job carbon cost between the two servers as the metric.
% A ratio higher than 1 means that the A10G server has a higher cost, and the V100 server will be better.
% As shown in Fig.~\ref{fig:sensitivity_fig1}, the higher carbon intensity and higher energy efficiency improvement would encourage the provisioner towards the new machines.
% For example, when the CI equals 0.9\,kgCO$_2$e$/$kWh, which is the carbon intensity of fuel energy\cite{ollivier2022sustainable}, the A10G only needs to have a 1.2$\times$ higher energy efficiency to flip the design.
% When the CI equals to 0.4\,kgCO$_2$e$/$kWh, which is the average carbon intensity of the state of Ohio in the US in 2022 \cite{electricitymaps}, the A10G needs to reach an improvement of 1.8$\times$ in energy efficiency to reach the crossover.

% \subsection{Generality Among Different Applications}
Fig.~\ref{fig:multi_app_fig1} shows the comparison between year 4 V100 servers and year 1 A10G servers under different depreciation models, carbon intensities, and applications. 
In this comparison, we still use the ratio of per-job carbon cost between V100 and A10G.
The two servers are set to have the same required throughput such that the V100 server runs at 60\% utilization.
In our profiling, persimmon-8b fails on V100 insufficient memory.

When the linear model is used, the per-job carbon cost will rely more on the operational carbon cost. Thus in most applications, A10G servers have a lower cost.
However, in the LLM workloads, the V100 outperforms A10G.
In these models, though A10G has a slightly higher throughput, its energy efficiency is even lower than V100.
When it comes to other depreciation models, the difference in the embodied cost will have a larger impact.
For all the eight applications profiled, the V100 server has a lower per-job carbon cost under the DB model.
Other depreciation models distribute the embodied carbon cost less aggressively, thus the per-job carbon cost between old and new servers will be closer.
% Moreover, in the application resnet 101, where the A10G has 1.52x higher energy efficiency, the A10G outperforms V100 when carbon intensity reaches 0.25\,kgCO$_2$e$/$kWh under the DDB model, and when the carbon intensity reaches 0.15\,kgCO$_2$e$/$kWh under the SYD model.

Fig.~\ref{fig:multi_app_fig2} shows the comparison results between the year 4 A10G and year 1 RTX5000 servers.
The results generally follow the trend explained above.
In addition, for all 9 applications, the A10G server outperforms the RTX5000 when carbon intensity is 0.05\,kgCO$_2$e$/$kWh even under the linear model.
This is because the RTX5000 has a 1.55$\times$ higher total embodied cost than A10G.
As a result, it needs more gain in operational cost to balance the overall cost.
% As a result, the RTX5000 needs to gain more operational carbon to balance the overall cost.
% For example, when running the ResNet-50 workloads, the RTX5000 will have a 1.98$\times$ higher throughput and 1.48$\times$ higher energy efficiency.
% However, for 1 million inferences, the carbon cost of RTX5000 and A10G are 7.4 and 4.8\,gCO$_2$e under the linear depreciation model, whereas the per-task energy energy consumption is 0.08 and 0.106\,kWh.
% As a result, the RTX5000 will not be better than the A10G until the carbon intensity increases to 0.1\,kgCO$_2$e$/$kWh so that the gain in operational carbon can balance .
\begin{yellowbox}
{\textbf{Observation.}} 
The effect of the depreciation model can flip the design choice between new and old devices in most cases.
\end{yellowbox} 

\section{Multi-device Clusters Discussion}

in reality, a data center will be a continuum of machines of different ages with different capabilities.  In this section we explore a data center with different classes and ages of machines.  In particular, we look at a replacement (provisioning) policy that supports lifetimes of 3 years, typical of data centers until recently~\cite{data-center-lifetime3} and a policy that supports lifetimes of 6 years, which is closer to the aspirational goal for more sustainable data centers that account for embodied carbon.  We see these are relative extreme points such that other configurations would lie in between.

% \subsection{Setup and Parameters}
We model a data center with six unique classes and ages of servers. 
The setup about the age and lifetime of servers follows the notion of replacement periods: the provisioner would replace a fixed portion of the oldest servers in the cluster per year, and the lifetime of all servers would be the same.
For a workload, we set a target throughput to model the total amount of work.  
We then distribute jobs to each server (server-class and age) following different scheduling policies.  
We set lower bounds of utilization for each device to 0.2 since even less efficient devices will be needed in bursty periods. 
We also set the upper bound to 0.9.
The carbon intensity is set to 0.025kgCO$_2$e$/$kWh, which is approximately the one in Ontario, Canada~\cite{ontario}.
We report two scheduling methods: (1) \textbf{Prefer New:} tries to allocate workloads to new devices until reaching the upper bound. (2) \textbf{Prefer Old:} Similar to prefer new, but prioritizes old devices.
These are greedy strategies to minimize carbon when operational or embodied carbon dominates, respectively.

% To understand the impact of carbon accounting on greedy scheduling we report two scheduling methods as follows:

% \noindent\textbf{Prefer New:} The scheduler will then try to allocate workloads to all new devices in a round-robin manner until reaching the upper bound of utilization. The remaining work will be allocated in round-robin fashion to older devices. This is a greedy strategy to minimize carbon per job when operational carbon dominates.

% \noindent\textbf{Prefer Old:} The scheduler is like Prefer New except it attempts to fulfill jobs on older devices instead.  This is a greedy strategy to minimize carbon per job when embodied carbon dominates.
\subsection{Impact of Different Depreciation Models}
\begin{figure}[tb]
\centering
\includegraphics[width=0.75\columnwidth]{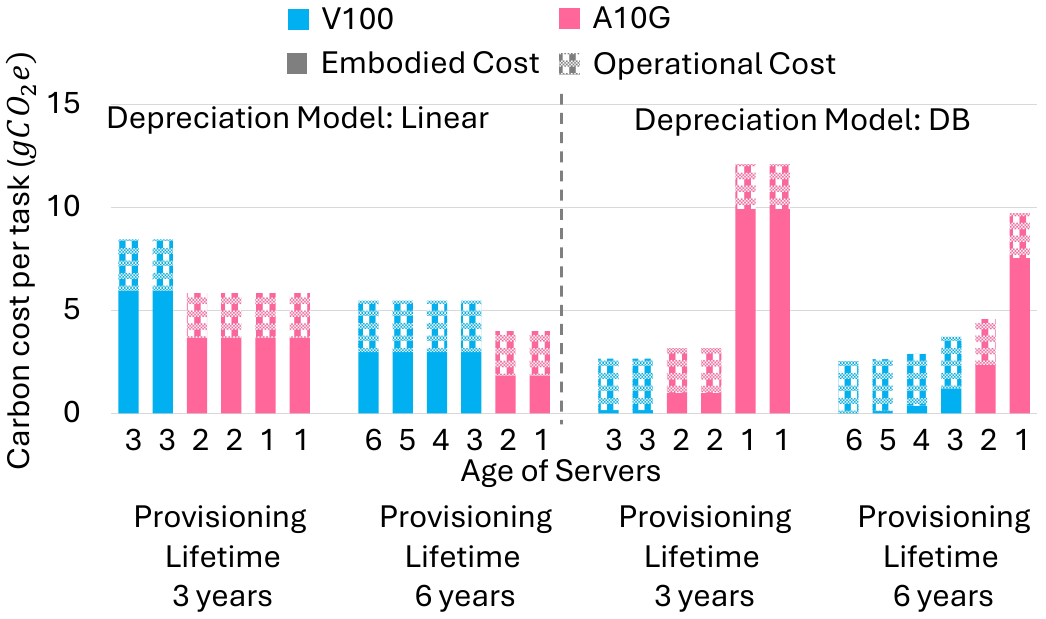}
\vspace{-12pt}
\caption{
    The per-job carbon cost of each server under different provisioning strategies and depreciation models in 2021.    
}
\label{fig:sche_knob1}
\vspace{-10pt}
\end{figure}

To analyze the impact of different depreciation models on job cost in a data center with provisioning and replacement cycles faster than hardware release, we show an example with six machine groups in Fig.~\ref{fig:sche_knob1}.
With linear depreciation, older hardware incurs a higher carbon cost per job.
With non-linear (DB) depreciation, job costs better reflect hardware age as newer hardware has a higher carbon cost, whereas V100 systems are relatively ``cheap".
Thus, a greedy scheduler favors running jobs on older hardware.
The average A10G job cost is over $2\times$ that of a V100 job for either lifetime.  

\begin{yellowbox}
{\textbf{Observation.}} 
Under linear depreciation models~\cite{SCI_project,switzer2023junkyard}, regardless of provisioning/replacement cadence, new hardware remains preferred.  Only non-linear depreciation allows a scheduler that optimizes jobs for carbon cost to prefer older hardware.
\end{yellowbox}

\subsection{Impact of Secondary Carbon Accounting}
\begin{figure}[tb]
\centering
\includegraphics[width=0.75\columnwidth]{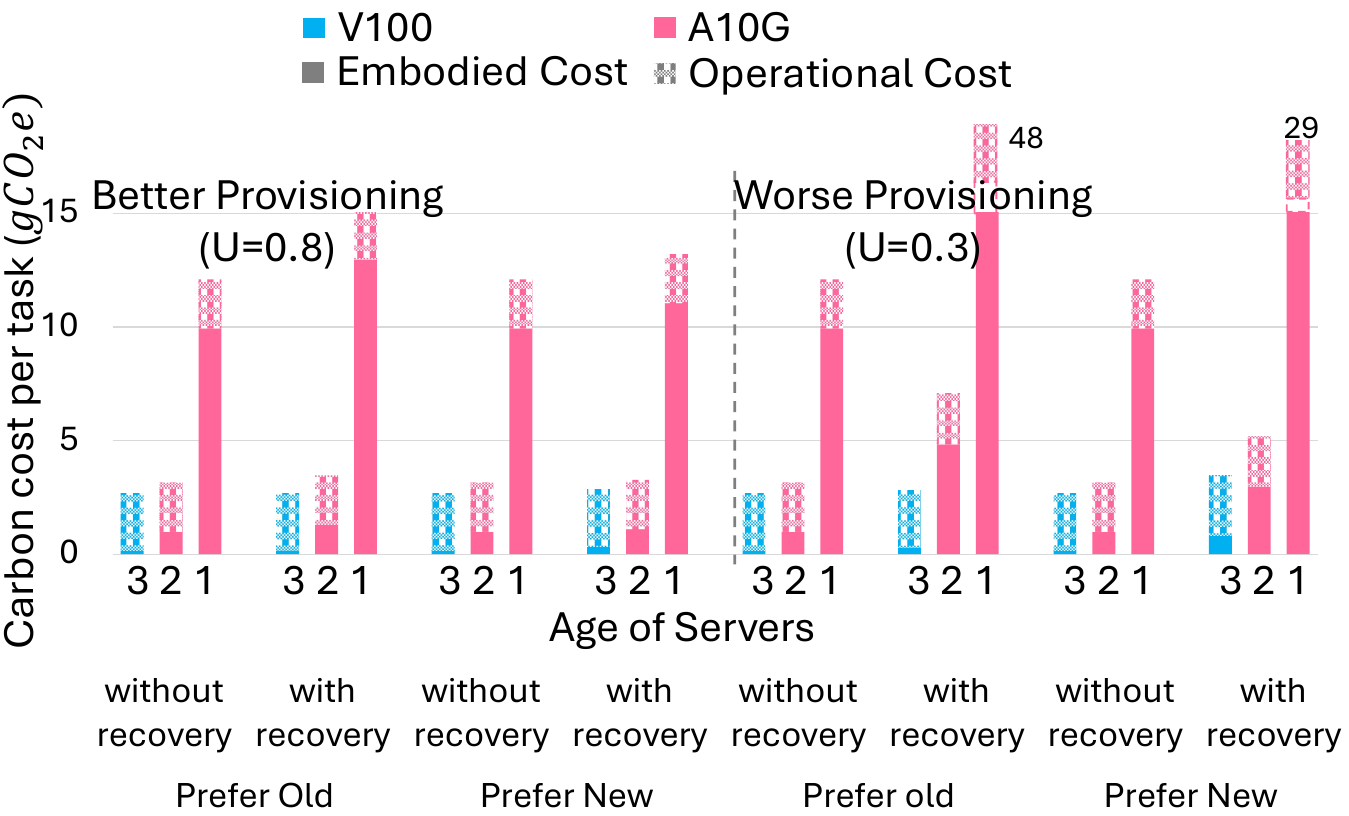}
\vspace{-12pt}
\caption{
    The per-job carbon cost of each server with or without secondary carbon recovery and under different throughput and scheduling policies in 2021. 
    % The DB depreciation model is used and the lifetime is set to 3 years..
}
\label{fig:sche_knob2}
\vspace{-15pt}
\end{figure}

Fig.~\ref{fig:sche_knob2} shows a 2020 snapshot of a data center with 3-year lifetime devices: two year-1 A10G groups, two year-2 V100 groups, and two year-3 V100 groups. 
We evaluate two utilizations, $U\in\{0.8,0.3\}$, representing better and worse provisioning, respectively.
At $U=0.8$, per-job carbon costs are similar across setups. Under secondary carbon accounting, ``prefer old" is slightly worse than ``prefer new" because new servers are a sunk embodied carbon cost and should be used as much as possible. 
However, improved provisioning outweighs scheduling differences, as using fewer servers saves substantial embodied carbon. 
Without secondary carbon, low utilization yields a per-job cost similar to high utilization. 
With secondary carbon included, low utilization significantly increases per-job cost. 
Thus, secondary carbon reveals a tunable gap for optimizing provisioning. 
At low utilization, ``prefer old" further raises the average cost because underused A10G servers produce high embodied-carbon-related secondary emissions. 
This trend correctly rewards good provisioning.

\begin{yellowbox}
{\textbf{Observation.}} 
Only when the secondary carbon cost is considered will the per-job carbon cost of a data center reflect better or worse provisioning scenarios.
\end{yellowbox}

\section{Conclusion}

In this paper, we explore non-linear depreciation models and show how they encourage the use of older hardware.
We also examine how secondary carbon accounting promotes higher server utilization and sustainable provisioning.
Linear depreciation can treat jobs on newer hardware as 25\% less carbon than those on older hardware.
In contrast, D/DB depreciation allows older hardware to recoup less carbon (jobs are $\geq 2\times$ cheaper) while saving embodied carbon during provisioning.
We further show that carbon recovery aligns with depreciation models and secondary carbon accounting.

%We propose two methodologies, the carbon depreciation models and the secondary carbon recovery, to address the importance of embodied carbon cost in data center provisioning.
%Our methodology encourages provisioning the right amount of servers and keeping the high utilization in clusters.
%We present the study at single-server and cluster level with various relevant workloads to demonstrate the effectiveness of the methodologies.

\smallskip
{\small
{\noindent\textbf{ACKNOWLEDGEMENTS --}} This work was partially supported by Brown University New Faculty Start-up Grant, DOE award DE-SC0026344,
NSF awards 
%\#2140346, % Dong's awards
%\#2231523, % Dong's awards
\#2348306, % Zhou's NSF PPOSS
%\#2441179, % Dong's awards
\#2511445, % Zhou's NSF MRI
\#2518375, % Zhou's NSF POSE
\#2536952, % Zhou's NSF DESC
\#2544032. % Zhou's NSF CAREER
We thank AMD for the hardware and software donations.
P. Zhou has a financial interest in Shanghai Suikun.

\balance
\bibliographystyle{ACM-Reference-Format}
% \bibliography{reference}
\bibliography{references_hpca25}

\end{document}